\documentclass[sigconf,nonacm]{acmart}

\usepackage{xcolor}
\usepackage{xspace}
\definecolor{bluegreen}{RGB}{0,153,153}
\definecolor{bluepurple}{RGB}{102,0,204}
\definecolor{bluered}{RGB}{204,0,102}

\usepackage{comment}

\AtBeginDocument{%
  \providecommand\BibTeX{{%
    \normalfont B\kern-0.5em{\scshape i\kern-0.25em b}\kern-0.8em\TeX}}}

\begin{document}

\title{An Annotated Dataset of Stack Overflow Post Edits}

\author{Sebastian Baltes}
\email{sebastian.baltes@adelaide.edu.au}
\orcid{0000-0002-2442-7522}
\affiliation{%
  \institution{University of Adelaide}
  \city{Adelaide}
  \state{South Australia}
  \postcode{5005}
  \country{Australia}
}

\author{Markus Wagner}
\email{markus.wagner@adelaide.edu.au}
\orcid{0000-0002-3124-0061}
\affiliation{%
  \institution{University of Adelaide}
  \city{Adelaide}
  \state{South Australia}
  \postcode{5005}
  \country{Australia}
}

\renewcommand{\shortauthors}{Baltes and Wagner}

\begin{abstract}

To improve software engineering, software repositories have been mined for code snippets and bug fixes. 
Typically, this mining takes place at the level of files or commits. 
To be able to dig deeper and to extract insights at a higher resolution, we hereby present an annotated dataset that contains over 7 million edits of code and text on Stack Overflow. 
Our preliminary study indicates that these edits might be a treasure trove for mining information about fine-grained patches, e.g., for the optimisation of non-functional properties.


%

\end{abstract}



\maketitle


\sloppy

\section{Motivation}

Data-Driven Search-Based Software Engineering (DSE)~\cite{Nair:2018:DSS:3196398.3196442} -- and the recent generalisation Data Mining Algorithms Using/Used-by optimisers (DUO)~\cite{agrawal2018better} -- 
combine insights from Mining Software Repositories (MSR) and Search-based Software Engineering (SBSE). While MSR formulates software engineering problems as data mining problems, SBSE reformulates SE problems as optimisation problems and use meta-heuristic algorithms to solve them. Both MSR and SBSE share the common goal of providing insights to improve software engineering. 
In this present paper, we suggest to improve software engineering -- in particular the search for code and text patches -- by mining the edit histories of Stack Overflow posts. 

We are, of course, not the first to propose to mine corpora for patches. In particular in the field of automated program repair~\cite{GouesPR19apr}, this has been a popular approach. For example, development histories of Eclipse JDT have been mined to find bug-fixing patches~\cite{Kim2013patchesfromhumans}, and so have been GitHub projects~\cite{Le2016historydriven,Long2017codetransforms}, but again for fixing bugs. 
Moving on from bug fixing to the optimisation of non-functional properties, Petke~\cite{Petke2017newops} suggested in 2017:
\begin{quote}
``[...] to mine changes made by software developers [...] with particular focus on improvement of the software property of interest, such as runtime efficiency. The results can then be used to devise new mutation operators in the form of templates. [...]''    
\end{quote}
Interestingly, we are not aware of any work towards this, and one reason for this might be the scarcity of data. 
Having said this, there has been the work of Moura et al.~\cite{Moura2015mineEnergyCommits}, who focused on mining energy-saving commits from GitHub that ``had the explicit intention of saving energy''. They manually interpreted 290 commits and presented along with two other code transformations for very specialised cases that might be translatable into patches for automatic search.
While said article has been cited 28 times at the time of writing, none of the citing works and none of the other works that we are aware of had a broader focus on non-functional properties in general.


With this article, we present a dataset of annotated Stack Overflow post edits for future extraction of patches.
We are of the opinion that the edits of Stack Overflow posts are by nature more fine-grained than, e.g., commits in Github repositories, and hence they are more amenable to the extraction of patches that can result in new mutation operators -- we would go as far to saying that this is because Stack Overflow post edits are less formal (due to the forum-like style of the platform) than a software repository commit where each commit is, e.g., expected to fix a bug or to extend functionality.\footnote{As anecdotal evidence, we point at the two discussions \url{https://softwareengineering.stackexchange.com/q/74764/} and \url{https://stackoverflow.com/q/107264} which have been viewed over 70k at the time of writing; accessed on 14 April 2020.}

Before we can start to extract patches, we first need to set the expectations by characterising the dataset. 
As we have 
a particular interest in code-optimisation, e.g., for the purpose of genetic improvement of software~\cite{Langdon2015gi,Petke2018gisurvey},  this greatly influences our research questions: 
\begin{enumerate}
    \item \textbf{RQ1:} Which aspects do Stack Overflow users mention in their edit comments? 
    \item \textbf{RQ2:} Which non-functional properties do users reference in edit comments?
\end{enumerate}



\section{Stack Overflow Post Edits}

\begin{figure}
    \centering
    \includegraphics[width=\linewidth,trim=230 85 50 40,clip]{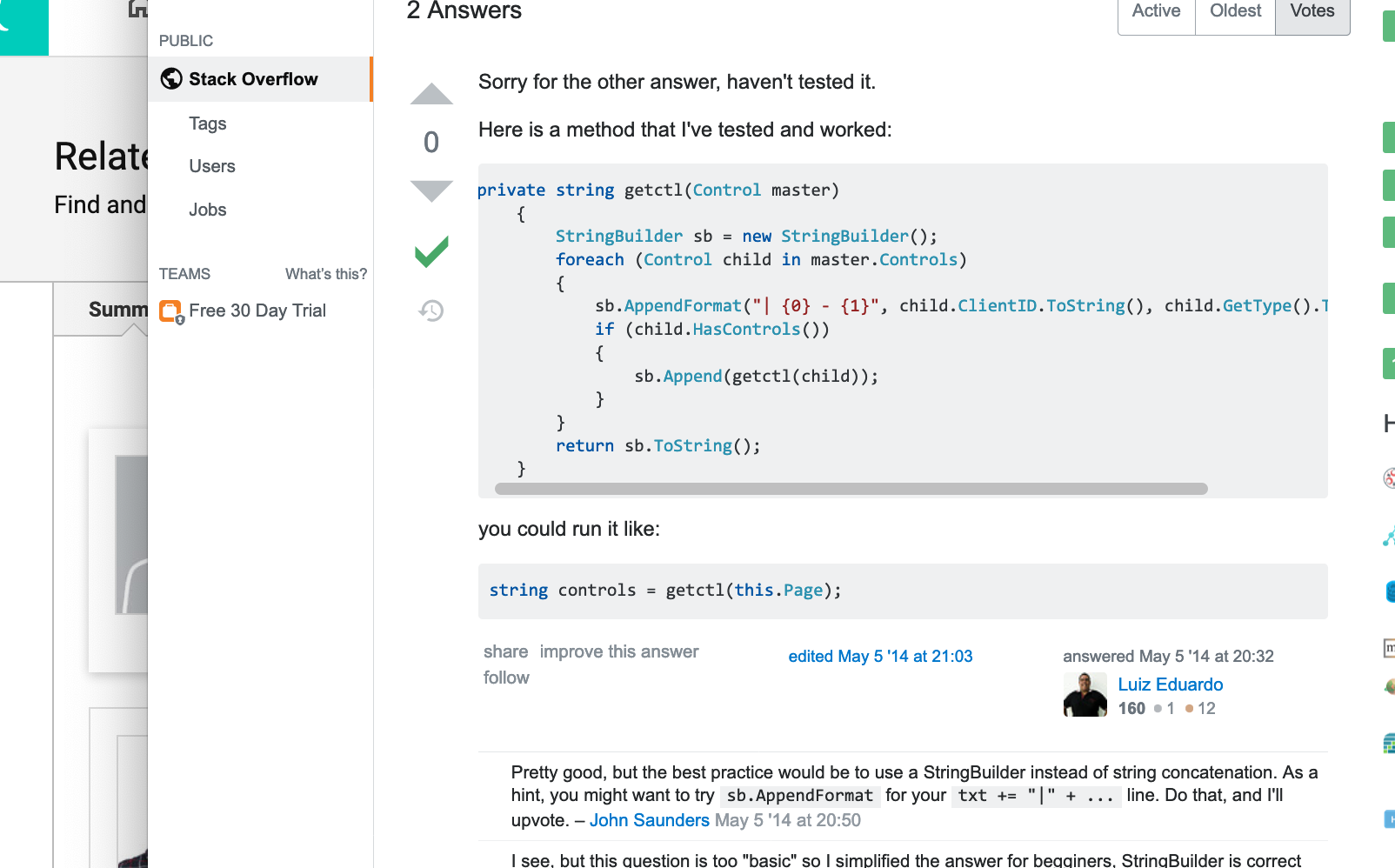}
    \caption{Example of a Stack Overflow post that contains text and code blocks and that has been edited after its initial publication. Source: \url{https://stackoverflow.com/a/23480549}, accessed on 17 April 2020.}
    \label{fig:soexample}
\end{figure}

To derive Stack Overflow code edits, we utilised the \emph{SOTorrent} dataset that we developed and maintain~\cite{DBLP:conf/msr/BaltesDT008}.
Based on version \emph{2020-01-24} of the dataset we retrieved all 7,459,778 post edits where the user provided an (optional) description of the edit.
Those edits, which are not limited to a particular programming language, are the foundation of our dataset~\cite{baltes_sebastian_2020_3754159}.
Of all edits, 1,305,323 (17.5\%) modified only a code block, 4,792,777 (64.2\%) only a text block, and 1,361,678 (18.3\%) both text and code blocks of a particular Stack Overflow post -- Figure~\ref{fig:soexample} shows an example of a typical post.

As a next step, we normalised the edit messages by converting all characters to lower case and by replacing all whitespace sequences with a single space character.
This yielded 3,291,268 unique edit messages. 

\begin{figure}[h]
  \centering
  \includegraphics[width=\linewidth, trim=0.0in 0.2in 0.0in 0.4in]{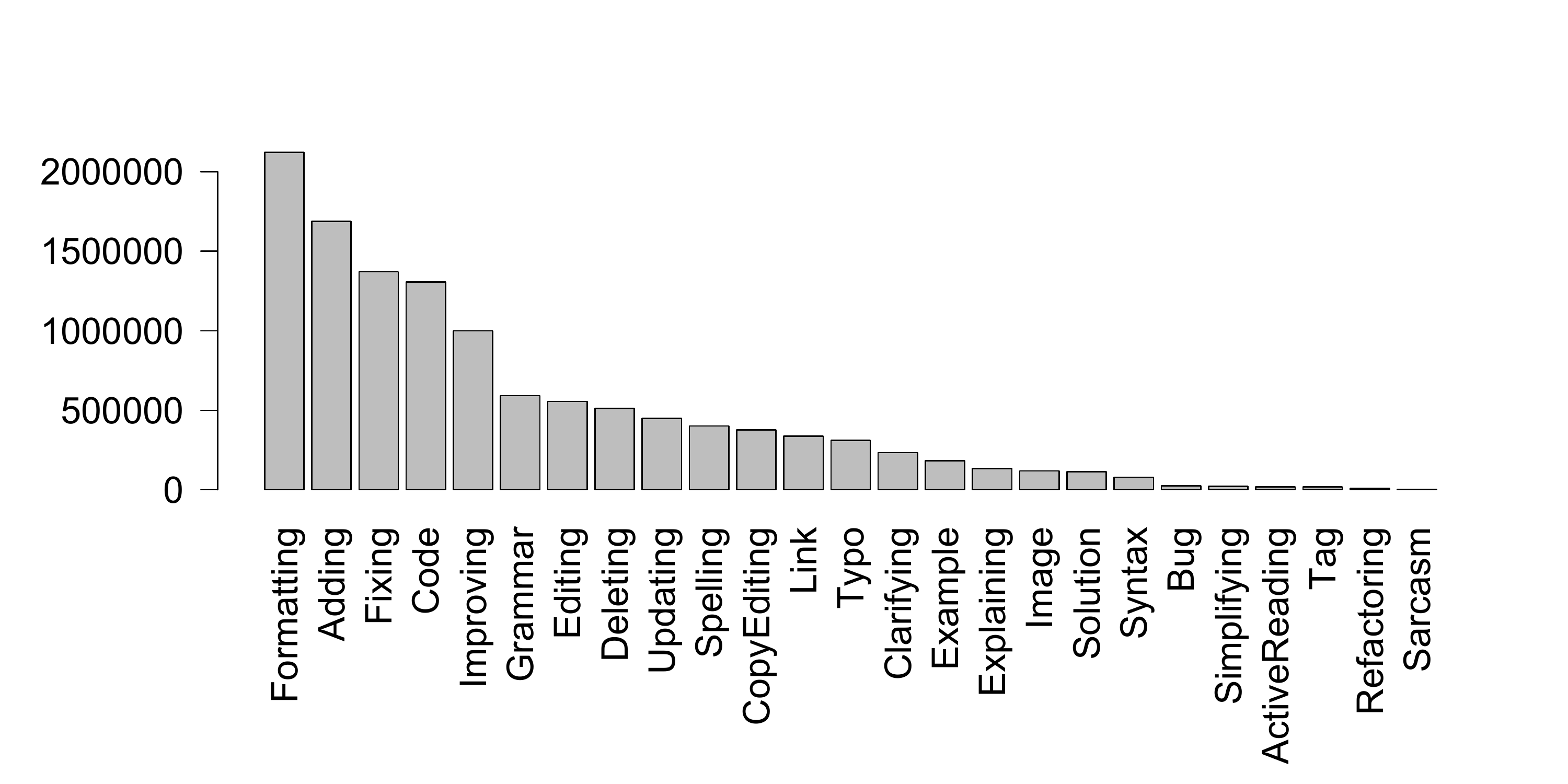} 
  \caption{Number of edits assigned to the 25 categories we iteratively derived (n=6,704,541).}
  \label{fig:uniquecomments}
\end{figure}

We then ranked the edit messages according to their frequency.
Starting with the most frequent messages, we manually extracted characteristic keywords to build regular expressions matching similar messages.
We stopped the manual analysis as soon as we were able to cluster all messages with at least 1,000 occurrences.
After this process, we were able to assign edit messages to 25 categories using customised regular expressions.
Thirteen categories were related to actions that the user performed (\emph{adding, updating, deleting, fixing, improving, clarifying, simplifying, explaining, editing, copy-editing, active reading, refactoring}), eleven were nouns related to the target of the edit (\emph{formatting, typo, grammar, spelling, code, bug, link, image, example, syntax, solution, tag}).
One category was more on a meta-level and captures \emph{sarcasm} in edit messages.
To provide an example, we present the final regular expression that we used to match edits \emph{adding} a certain information:
\begin{verbatim}
.*\\b((add|expand|more|extend)[a-z0-9_-]*).*
\end{verbatim}
Please note that one edit can belong to more than one of these edit categories.
Overall, we were able at assign 6,704,541 of the 7,459,778 edits (89.9\%) to at least one category (see Figure~\ref{fig:uniquecomments}). 
We provide our retrieval and analysis scripts as part of our dataset~\cite{DBLP:conf/msr/BaltesDT008}.

Figure~\ref{fig:uniquecomments} shows the results of the above-mentioned annotation process, thus answering \textbf{RQ1}.
Most post edits are concerned with \emph{formatting} or \emph{additions}.
Many edits referred to \emph{fixing} and to \emph{code}, which indicates that our dataset is indeed suitable for mining patches.
While the term \emph{bug} was less frequently mentioned, it still occurred in 24,775 edit messages.

\begin{figure}[h]
  \centering
  \includegraphics[width=\linewidth, trim=0.0in 0.2in 0.0in 0.4in]{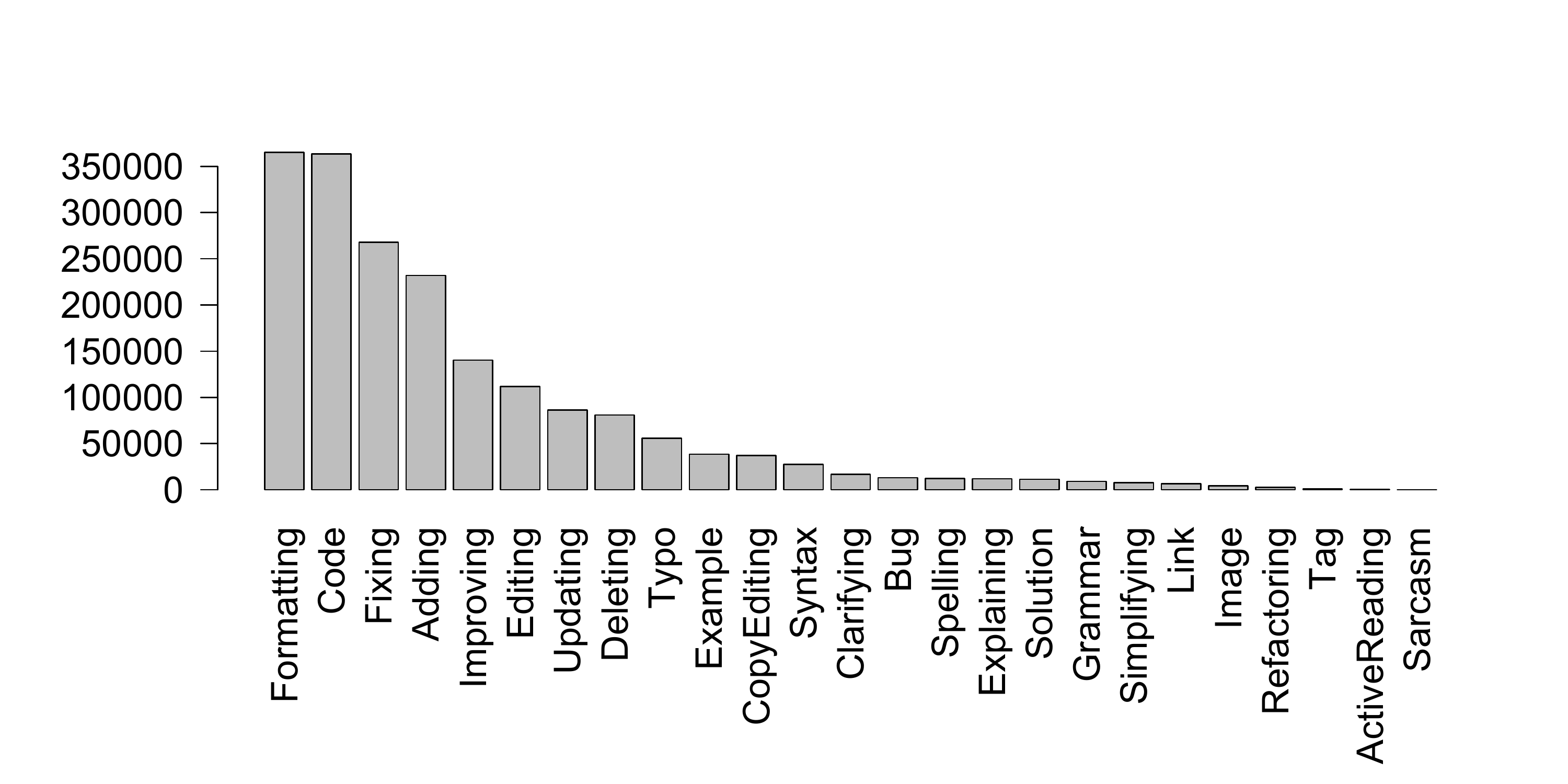} 
  \caption{Number of code edits assigned to the 25 categories we iteratively derived (n= 933,340).}
  \label{fig:uniquecommentsCODEonly}
\end{figure}

In the following, to explore the potential for patch-extraction, we focus on post edits that only edited code blocks, because this allows us to unambiguously link the edit message to the code edit.
Of the 1,305,323 edits that only modified code blocks, 933,340 (71.5\%) were assigned to at least one of the 25 categories.
Figure~\ref{fig:uniquecommentsCODEonly} shows the corresponding distribution.
As we can see, about one third of these edits are about \emph{formatting} and \emph{code}, but \emph{fixing} and \emph{adding} were also frequently mentioned.
Overall, the edits that are interesting for us are the hundreds of thousands of edits that are concerned with \emph{fixing}, \emph{editing}, \emph{updating}, or \emph{improving}.

Next, we dig a little deeper. In Table~\ref{tab:codeonlyeditsTop10}, we list the most frequent pairs of tags (of edits that were assigned to at least two categories). While a large proportion of the post edits is about the \emph{formatting} of \emph{code}, over 230,000 edits possibly target the improvement of a code aspect of a post, i.e., by \emph{fixing}, \emph{adding}, \emph{improving}, \emph{editing}, \emph{updating}, or \emph{deleting} code.

\begin{table}[]
    \centering
    \begin{tabular}{lr}
    \toprule
\textbf{Pair} & \textbf{Count} \\\midrule
formatting,	code&	152,721 \\
improving,	formatting	&98,339 \\
fixing,	code&	76,026 \\
fixing,	formatting&	65,544 \\
adding,	code&	50,795\\
fixing, typo	&	32,711\\
improving, code&	31,463\\
editing, code&	28,844\\
updating, code&	24,910\\
deleting, code&	20,106\\\bottomrule
    \end{tabular}
    \vspace{2mm}
    \caption{Top 10 pairs of tags (pairs ordered only for presentation purposes).}
    \label{tab:codeonlyeditsTop10}
\end{table}

Finally, we focus on edits targeting non-functional properties, as this appears to be an area that is currently under-explored in the literature. 
While non-functional properties were not among the most frequently mentioned terms that formed the 25 categories, we want to show that our dataset nevertheless contains edits related to such properties.
We derived four categories capturing non-functional properties based on our own judgement and information taken from the Wikipedia page on non-functional requirements.\footnote{\url{https://en.wikipedia.org/w/index.php?title=Non-functional_requirement&oldid=947189406}, accessed on 13 April 2020}
We again built custom regular expressions to match edits related to these categories (see scripts attached to our dataset~\cite{DBLP:conf/msr/BaltesDT008}).
Table~\ref{tab:nf49wiki11manual} shows the results of applying the regular expressions to the edit messages of all code-only edits, thus answering \textbf{RQ2}. 
We found a few thousand edits that appear to target non-functional properties such as \emph{performance} and \emph{memory}. 
Interestingly, the small number of edits targeting \emph{energy} is in line with the small number of commits that the aforementioned research~\cite{Moura2015mineEnergyCommits} has been able to find.

\begin{table}[]
    \centering
    \begin{tabular}{lr}
    \toprule
\textbf{Property} & \textbf{Count} \\\midrule
performance&	2,658 \\
size	& 2,284 \\
memory&	1,084 \\
energy&	10 \\\bottomrule
    \end{tabular}
    \vspace{2mm}
    \caption{Number of code edits where the user mentioned one of the four non-functional properties we have considered (n=7,024).}
    \label{tab:nf49wiki11manual}
\end{table}

\section{Examples}

Our dataset can either be downloaded from Zenodo~\cite{DBLP:conf/msr/BaltesDT008} as a CSV file or accessed via Google BigQuery.\footnote{\url{https://bigquery.cloud.google.com/table/sotorrent-org:2020_01_24_edits.PostEdits}}
The table \texttt{PostEdits}, which we provide as part of our dataset, gives researchers access to all 7,459,778 post edits extracted from the \emph{SOTorrent} dataset, identified by their \texttt{PostHistoryId}.
We further provide the edit messages and binary flags for the categories mentioned throughout this paper, which can be used to filter the edits.
The table can be joined with table \texttt{PostBlockVersion} of the \emph{SOTorrent} dataset to retrieve the content of the modified text and code blocks before and after the edits.
A corresponding query is attached to our dataset~\cite{DBLP:conf/msr/BaltesDT008}.


As a first investigation to explore the potential of our dataset, we have manually explored the subset of the code-block edits that we had tagged as being \emph{performance}-related. 
For this proof-of-concept, a total of 15 minutes was spent on the exploration of the edits and on the assessment of the respective Stack Overflow posts.
Among others, we have found the following edits:\footnote{structure: edit comment from the post editor (post ID): our interpretation}
\begin{enumerate}
\item \emph{``using john saunders tip for more performance''} (23481309): the edit replaced a String with a StringBuilder (see Figure~\ref{fig:soexample}).
\item \emph{``added debounce to improve performance when app scales''} (44000037): the edit added a JavaScript debounce function.
\item \emph{``evaluating x 0 first solves for type errors and gives better performance than if''} (19400435): the edit updated an if-statement -- interestingly, there is a brief discussion on the performance attached to this post.
\item \emph{``some small performance improvements always a good idea to have a fast primality test''} (8539774): the edit added a few hard-coded scenarios for a particular problem.
\item \emph{``Improved performance, by getting [...] outside the loop''} (11535593): the edit lifted code outside of a loop, which is an approach that is commonly taught in undergraduate courses.
\end{enumerate}

\section{Outlook}

The particular value of our dataset is that the post edits are most likely much smaller in scope and much more fine-grained than, e.g., commits in project repositories. With his hypothesis in mind, it might be possible to reveal insights on software engineering in practice at a higher resolution.
Moreover, our preliminary study indicates that the Stack Overflow edits might be a treasure trove for manually mining information about fine-grained code patches, e.g., for the optimisation of non-functional properties. 
Lastly, while our focus has been on code edits, we envision that our dataset -- which also contains text edits and their comments -- can be of use for mining other insights as well, including typical grammar fixes or frequent formatting improvements.

We are open for feedback from the community on potential improvements to the dataset and we are happy to provide support for researcher who want to use or adapt our data.
Researchers can, for example, add additional categories by adapting our regular-expression-based matching approach.

\bibliographystyle{ACM-Reference-Format}
\bibliography{geccogi20-soedits,dp21d1}

\end{document}